\documentclass[12pt]{article}
\newcommand{\rrangle}{{\rangle\!\rangle}}
\newcommand{\llangle}{{\langle\!\langle}}
\newcommand{\cC}{\mathcal{C}}
\newcommand{\cN}{\mathcal{N}}

\begin{document}

\begin{titlepage}
\thispagestyle{empty}
\begin{flushleft}
\hfill hep-th/0402107 \\
UT-04-06\hfill February, 2004 \\
\end{flushleft}

\vskip 1.5 cm
\bigskip

\begin{center}
\noindent{\LARGE Cardy states as idempotents}\\
{\LARGE of fusion ring in string field theory}

\renewcommand{\thefootnote}{\fnsymbol{footnote}}

\vskip 2cm
{\large
Isao~Kishimoto\footnote{e-mail address:
 ikishimo@hep-th.phys.s.u-tokyo.ac.jp} and
Yutaka~Matsuo\footnote{e-mail address:
 matsuo@phys.s.u-tokyo.ac.jp}}
\\
{\it
\noindent{ \bigskip }\\
Department of Physics, Faculty of Science, University of Tokyo \\
Hongo 7-3-1, Bunkyo-ku, Tokyo 113-0033, Japan\\
\noindent{ \smallskip }\\
}
\vskip 20mm
\bigskip
\end{center}
\begin{abstract}
With some assumptions, the algebra between
Ishibashi states in string field theory can be reduced 
to a commutative ring. {}From this viewpoint, 
Cardy states can be identified with its idempotents.
The algebra can be identified with a fusion ring for 
the rational conformal field theory and a group ring for  the orbifold.
This observation supports our previous observation
that boundary states satisfy a universal idempotency 
relation under closed string star product.
\end{abstract}
\end{titlepage}\vfill\setcounter{footnote}{0}
\renewcommand{\thefootnote}{\arabic{footnote}}

\newpage
\paragraph{Commutative rings}
In this short note, we consider projection operators
of two types of commutative rings
which are related to the conformal field theory.
The first one is the group ring 
$\mathbf{C}^{[\Gamma]}$ \cite{r:group} based on
a discrete finite group $\Gamma$. It is defined as a
vector space with $|\Gamma|$ dimensions.  
If we write its basis as $e_g$ ($g\in\Gamma$), generic elements
can be expanded as 
$\lambda=\sum_{g\in\Gamma} \lambda_g \, e_g$,
$ \lambda_g\in\mathbf{C}\,.$
It is equipped with a natural  product structure,
\begin{equation}
\label{group_ring}
\lambda\star \mu =\sum_{g,g'\in \Gamma} \lambda_g\mu_{g'} e_{gg'}
=\sum_{g\in \Gamma}\left(\sum_{g_1 g_2=g} \lambda_{g_1}\mu_{g_2}\right)e_g
\end{equation}
where $gg'$ is the product in the group $\Gamma$.

When the group $\Gamma$ is 
non-abelian, we combine
the basis which belong to the same conjugacy class.
Namely instead of treating the basis $e_g$ separately,
we take a combination $ e_i = \sum_{g\in\cC_i}e_g$
as the basis where $\cC_i$ is a conjugacy class.
 While the group ring becomes noncommutative,
the combined basis satisfies a commutative algebra,
\begin{equation}
\label{class_prod}
e_i\star e_j =\sum_{j}\cN_{ij}{}^k \,e_k\,.
\end{equation}
The structure constant $\cN_{ij}{}^k$ is non-negative integer and
can be written in terms of the characters of the irreducible representations
of $\Gamma$ as \cite{r:group},
\begin{equation}
\label{Verlinde2}
\cN_{ij}{}^k =\frac{1}{|\Gamma|}\sum_{\alpha\in \mbox{\scriptsize irreps.}}
\frac{r_i r_j \zeta_i^{(\alpha)} \zeta_j^{(\alpha)}\zeta_k^{(\alpha)*}}
{\zeta_1^{(\alpha)}}\,,
\end{equation}
where $\zeta_i^{(\alpha)}$ is the character of an irreducible
representation $\alpha$ for the elements in the conjugacy class $\cC_i$.  
$r_i$ is the number of elements in $\cC_i$. We take $\cC_1$ as the
conjugacy class that consists only of the identity element.
This formula is analogous to Verlinde formula (\ref{Verlinde})
in the following.

Another algebra in consideration
is the fusion ring \cite{Verlinde:sn,Kawai:wg}
 of rational conformal field theories 
(RCFT).
Suppose there exist $n$ primary fields $\phi_i$ ($i=1,\cdots,n$). The
fusion ring for RCFT consists of the $n$ dimensional vector space
with base $e_i$ ($i=1,\cdots,n$) together with
a product structure for the basis,
\begin{equation}
\label{fusion_ring}
\left(\sum_i \lambda_i e_i\right) \star \left(\sum_j \mu_j e_j\right)= \sum_{k}
\left(\sum_{ij} \lambda_i \mu_j N_{ij}{}^k\right) e_k\,,
\end{equation}
where $N_{ij}{}^k$ is Verlinde's fusion coefficient \cite{Verlinde:sn} 
defined by the modular transformation matrices,
\begin{equation}
\label{Verlinde}
N_{ij}{}^k=\sum_{l}\frac{S_{il} S_{jl}S^*_{kl}}{S_{1l}}\,. 
\end{equation}
$S$ is symmetric, unitary and satisfies $(S^2)_{ij}=\delta_{ji^*}$
where $i^*$ is a charge conjugation of $i$.

In both cases, the product $\star$ satisfies the associativity.  For the
group ring, this is obvious.  
For the fusion ring, it reduces to a property
of the fusion multiplicity,
\begin{equation}
\label{associativity}
\sum_l N_{ij}{}^l N_{lk}{}^m=\sum_l N_{jk}{}^l N_{il}{}^m\,,
\end{equation}
which can be proved directly from 
Eq.(\ref{Verlinde}) and the unitarity of
$S$.

\paragraph{Results from closed string field theory}
Our motivation to consider the commutative rings comes
from string field theory.
In our previous papers \cite{Kishimoto:2003ru,Kishimoto:2003yu},
we have derived properties 
of the boundary states with respect to the star product of closed string field theories 
\cite{Hata:1986kj, Zwiebach:1992ie}. 
\begin{enumerate}
\item Suppose we have two boundary states which satisfies,
\begin{equation}
\label{conformal_inv}
(L_n-\tilde L_{-n})|B_i\rangle =0\,,\quad (i=1,2)\,.
\end{equation}
Then a state which is created by HIKKO's star product 
\cite{Hata:1986kj}
which specifies interactions of closed strings
satisfies the same equation,
$(L_n-\tilde L_{-n})(|B_1\rangle * |B_2\rangle)=0$.
This claim is proved in the background independent fashion
and should be applied to any conformal field theories.
\item The second observation is the idempotency of
boundary states.  It was proved in flat background
that boundary states $|x^\perp,F\rangle$ of
D$p$-branes with transverse coordinates $x^\perp$ and flux $F$
satisfy
\begin{equation}
\label{idempotency}
|x^\perp,F\rangle_{\alpha_1}* |y^\perp,F\rangle_{\alpha_2}
 = [\cC(\alpha_1,\alpha_2)]^d
 \delta^{(d-p-1)}(x^\perp-y^\perp) |y^{\perp},F\rangle_{\alpha_1+ \alpha_2}\,,
\end{equation}
where $\alpha_i$ are parameters which specify the length of
overlapping closed strings at the vertex.  
The coefficient $\cC$ is given in
appendix B of Ref.~\cite{Kishimoto:2003yu},
$\cC(\alpha_1,\alpha_2)=T^{-1/8}|\alpha_1\alpha_2\alpha_3|^{1/24}\,$
 ($\alpha_3=-\alpha_1-\alpha_2$).
Here we write only the contribution from the matter part.\footnote{
\label{fn:1}
In the definition of $*$ for each boson, we absorbed
a factor $\mu^{1/12}$ where 
$\mu=e^{-\tau_0\sum_{r=1}^3 \alpha_r^{-1}}$, $
\tau_0=\sum_{r=1}^3 \alpha_r \log|\alpha_r|\,.
$
Including ghost sector, we should replace $[\cC(\alpha_1,\alpha_2)]^d$ with
$[\cC(\alpha_1,\alpha_2)]^{26-2} c_0^+$ in Eq.(\ref{idempotency}).}
$T$ is a constant cut-off parameter which we need to take $T\rightarrow 0$.
In section 3 of Ref.~\cite{Kishimoto:2003yu},
we also gave a geometrical picture  that explains the idempotency relation
for generic background CFT.
\end{enumerate}

\paragraph{Ishibashi and Cardy state}
A convenient basis of the states which satisfy 
Eq.(\ref{conformal_inv}) is Ishibashi
state \cite{Ishibashi:1988kg}. In a generic CFT, there exists
a state $|i\rrangle$  for each primary field $\phi_i$ 
which satisfies Eq.(\ref{conformal_inv}) and,
\begin{equation}
\label{Ishibashi}
\llangle i|\tilde q^{\frac{1}{2}(L_0+\tilde L_0-\frac{c}{12})}|j\rrangle =
\delta_{ij}\chi_j(\tilde q)\,,\quad
\tilde q=\exp(-2\pi i/\tau)\,,
\end{equation}
where $\chi_j$ is the character for the highest weight representation
associated with the primary field $\phi_j$.
Since Ishibashi states give the basis of the Hilbert space that
satisfies Eq.(\ref{conformal_inv}),
we can expand the star product between them,
\begin{equation}
\label{universal_rel}
|i\rrangle_{\alpha_1} * |j\rrangle_{\alpha_2}=
\sum_k \cC_{ij}{}^k(\alpha_1,\alpha_2)
\, |k\rrangle_{\alpha_1+\alpha_2}\,,
\end{equation}
where the coefficient $\cC_{ij}{}^k(\alpha_1,\alpha_2)$ is a
$c$-number. 
Here the $*$ product is HIKKO's one.
It is commutative
and  non-associative but satisfies Jacobi identity instead.

The conformal invariance of the boundary (\ref{conformal_inv}) is, 
actually, not
enough to define a consistent boundary state.  We have
to impose Cardy condition, in order to have well-defined
open string sector \cite{Cardy:ir} 
(see also Refs.~\cite{Behrend:1999bn, Billo:2000yb} for reviews).  
We take a linear combination of Ishibashi states,
$|\alpha\rangle=\sum_i \psi^i_\alpha |i\rrangle$, and calculate
the inner product between them.  After we perform a modular transformation,
\begin{equation}
\label{Cardy_cond}
\langle \alpha | \tilde q^{\frac{1}{2}(L_0+\tilde L_0 -\frac{c}{12})}
|\beta\rangle
=\sum_{i} (\psi_\alpha^i)^* \psi_\beta^i \chi_i(\tilde q)
=\sum_{j}n_{\alpha\beta}{}^j \, \chi_j(q)\,,\quad
n_{\alpha\beta}{}^j:=\sum_i (\psi_\alpha^i)^* \psi_\beta^i S_{ji}^*\,,
\end{equation}
where $\chi_i(q)=S_{ij}\chi_j(\tilde{q}),~(q=e^{2\pi i \tau})$.
In order to have a well-defined open string Hilbert space, the coefficient
$n_{\alpha\beta}{}^j$ must be non-negative integer.  This gives a
set of quadratic constraints for the coefficients $\psi^i_\alpha$.
A famous family of solutions found by Cardy is
\begin{equation}
\label{Cardy_sol}
\psi^i_\alpha = \frac{S_{\alpha i}}{\sqrt{S_{1i}}}\,.
\end{equation}
With this choice, the multiplicities $n_{\alpha^*\beta}{}^j$ 
coincides with the
fusion rule coefficient $N_{\alpha\beta}{}^j$ in Eq.(\ref{Verlinde}).
For a systematic study of other choices of $\psi$, see for example,
Ref.~\cite{Behrend:1999bn}.

The purpose of this note is to study if
 the idempotency relation is consistent 
with Cardy condition (\ref{Cardy_cond}).
While both equations are quadratic, they look rather different.
To compare the two conditions, we first
solve the idempotency relation
using Eq.(\ref{universal_rel})
and compare them with the solutions of Cardy condition.
As we see below, they are essentially the same.

\paragraph{Reduced SFT algebra for Ishibashi state}
In the following, we conjecture
 that the structure constant $\cC_{ij}{}^k(\alpha_1,\alpha_2)$
 in Eq.(\ref{universal_rel})
will take a factorized form $[\cC(\alpha_1,\alpha_2)]^c\, R_{ij}{}^k$
where $R_{ij}{}^k$ is a constant which is independent of $\alpha_i$
and $c$ is the central charge of the conformal field theory.

This assumption was explicitly proved for the flat background
\cite{Kishimoto:2003ru,Kishimoto:2003yu}
where Ishibashi state is the momentum eigenstate in the transverse
direction.  By an explicit computation in oscillator formulation, we found
$
|p^\perp_1\rrangle_{\alpha_1} * |p^\perp_2\rrangle_{\alpha_2} =
[\cC(\alpha_1,\alpha_2)]^d \, |p_1^\perp + p_2^\perp\rrangle_{\alpha_1
+\alpha_2}
$. 
As a  more nontrivial background, we have studied numerically the
${\mathbf Z}_2$ orbifold. We have confirmed the factorization while we need to
change the definition of the factor $\mu$ which appeared
in footnote (\ref{fn:1}) in the string vertex
which involves the twisted sectors.
The detail will be published elsewhere.

{}From the constant part of
the structure constant, one can define an algebra 
$e_i \star e_j=\sum_k R_{ij}{}^k e_k$ 
which is independent of  $\alpha_i$
where $e_i$ corresponds to $|i\rrangle$ in the reduced algebra.
Since HIKKO's product $*$ is commutative,
the reduced product $\star$ is commutative. 
Moreover we assume that the reduced product $\star$ is associative
for Ishibashi states which can be guessed from the associativity
in the flat background.
In the following, we call the simplified algebra as {\em
reduced SFT algebra}.

In the following, we fix $R_{ij}{}^k$
for two CFT models.
The first one is the orbifold $M/\Gamma$ with the discrete group $\Gamma$
(see, for example,  a review paper \cite{Billo:2000yb} and references therein).
In this case, the Ishibashi state $|g\rrangle$ has a label $g\in \Gamma$ which 
specifies the twisted boundary condition,
\begin{equation}
(X(\sigma+2\pi)-g\cdot X(\sigma))|g\rrangle=0\,.
\end{equation}
In this note, we consider a situation where there
is only one boundary state 
 which satisfies Eq.(\ref{conformal_inv})
for each twisted sector
for simplicity.
The star product $|g\rrangle * |g'\rrangle$ belongs to
the twisted sector of $gg'$.
The reduced algebra therefore should be written in the form
$e_g \star e_{g'}=R(g,g')e_{gg'}$.
The associativity of the $\star$ product imposes a condition
on the coefficient $R(g,g')$,\footnote{This condition is also imposed by
Jacobi identity for HIKKO's $*$ product without assuming the
associativity.}
\begin{equation}
\label{two-cycle}
R(gh,k)R(g,h)=R(g,hk)R(h,k)\,.
\end{equation}
A change of the normalization of the basis
 $e_g \rightarrow \beta(g) e_g$
will change the above coefficient:
 $R(g,h)\rightarrow \frac{\beta(gh)}{\beta(g)\beta(h)}R(g,h)$\,.
The nontrivial solutions of Eq.(\ref{two-cycle}) with this
 identification are parametrized by the cohomology group $H^2(\Gamma,
 U(1))$.  This is the discrete torsion factor
in the orbifold models \cite{Vafa:1986wx}. 
In this note, we consider the
situation where there is no such cohomology
 where
one can consistently 
put $R(g,h)=1$ up to the normalization factor of the boundary state.
The reduced SFT algebra is identical to group ring 
$\mathbf{C}^{[\Gamma]}$ in  Eq.(\ref{group_ring}).
When $\Gamma$ is non-commutative, it is known that we have
to combine the Ishibashi states of the same conjugacy class.
The reduced SFT algebra is thus the same as Eq.(\ref{class_prod}).

Our second example is the rational conformal field theory where we have
only finite number of primary states. We need to impose on the coefficient
$R_{ij}{}^k$ a few constraints (i) if the coefficient $N_{ij}{}^k$ vanishes
for a set ($ijk$), the corresponding $R_{ij}{}^k$ should also vanish
(ii) it needs to satisfy the associativity condition similar to 
Eq.(\ref{associativity}).
At this moment, we do not know the analogue of the cohomology group
which would fix $R_{ij}{}^k$ up to the normalization constant
as in the group ring case.  However, since $N_{ij}{}^k$ itself
explicitly solves them, we will focus on the fusion ring
(\ref{fusion_ring}) without further 
verification from the string field theory.
The outcome is, as we will see, natural and it supports
this assumption.

To summarize, up to some ambiguities, we regard the commutative rings
(\ref{group_ring}), (\ref{fusion_ring}) as the reduced SFT algebras of the
Ishibashi states. 
We note that if a projector of the reduced algebra is
found, for example in the form $\sum_i \lambda_i \,e_i$, 
the same combination $\sum_i \lambda_i |i\rrangle$ becomes
an idempotent of the full string field algebra.

\paragraph{A simple example of  projectors in the group ring}
We consider the projection operators for the star product $\star$ for these
commutative rings.  Before writing
down somewhat abstract formula from the beginning,
we start from a simple example which illuminates the basic structure.
We consider the case $\Gamma=\mathbf{Z}_3$. We write its elements
as $\left\{1,g,g^2\right\}$ where $g^3=1$. We can solve the algebraic
equation $\lambda\star\lambda=\lambda$ with $\lambda=\alpha_0 e_1
+\alpha_1 e_g +\alpha_2 e_{g^2}$.
It has three quadratic equations for three unknown variables.
It therefore has $2^3=8$ solutions,
\begin{eqnarray}
(\alpha_0,\alpha_1,\alpha_2) & = & (0,0,0), \nonumber\\
&& \frac{1}{3}(1,1,1),~~
\frac{1}{3}(1,\omega,\omega^2),~~
\frac{1}{3}(1,\omega^2,\omega),\nonumber\\
&& \frac{1}{3}(2,-1,-1),~~\frac{1}{3}(2,-\omega,-\omega^2),~~
\frac{1}{3}(2,-\omega^2,-\omega),\nonumber\\
 &  & (1,0,0), \nonumber
\end{eqnarray} 
with $\omega=e^{2\pi i/3}$.
The first line is a trivial solution.  We will identify the three
solutions in the second line with the fractional D-branes 
in $\mathbf{Z}_3$ orbifold.
They are written in terms of Ishibashi states as,
\begin{equation}
\label{fractional_D}
|\alpha\rangle=\alpha_0 |1\rrangle +\alpha_1 |g\rrangle
 +\alpha_2|g^2\rrangle
\end{equation}
where $|1\rrangle$ (resp. $|g\rrangle$) is the boundary state
for the untwisted ($g$-twisted) sector.
After this interpretation, the solutions in the third line correspond to
the combinations of two different fractional D-branes and the fourth
line corresponds to the non-fractional D-brane that corresponds to the
regular representation of $\mathbf{Z}_3$.

\paragraph{General projectors of group (fusion) ring and Cardy states}
We write down generic idempotents of the commutative rings
and compare them with Cardy states.
As we have seen in $\mathbf{Z}_3$ example,
there exist $2^n$ solutions to the idempotency relation
where $n$ is the number of the conjugacy classes (for group ring)
or that of the primary fields (for RCFT).
In both cases, the explicit form of the idempotents is known.
A novelty here is the comparison with the Cardy states.

For the group ring, there exists an idempotent for each
irreducible representation of $\Gamma$.
\begin{equation}
P^{(\alpha)}=\frac{d_\alpha}{|\Gamma|}\sum_{i\in \mathrm{Class}}
\zeta^{(\alpha)}_i \,e_i\,,
\end{equation}
where $d_\alpha=\zeta^{(\alpha)}_1$ is the dimension of the irreducible
representation \cite{r:group}.
For the fusion ring
\cite{Kawai:wg},
\begin{equation}
P^{(\alpha)}=S^*_{1 \alpha}\sum_{i\in\mathrm{Primary}} S_{i \alpha} e_i\,.
\end{equation}
In both cases, they  satisfy the idempotency relation,
$P^{(\alpha)}\star P^{(\beta)}=\delta_{\alpha\beta} P^{(\beta)}$. 
Note that we use the charge conjugacy symmetry $N_{i^*j^*}{}^{k^*}
=N_{ij}{}^k$ in the proof.
Because of the orthogonality, $2^n$ general solutions are immediately obtained,
$P=\sum_\alpha \epsilon_\alpha P^{(\alpha)}$
where 
$\epsilon_\alpha=0\mbox{ or }1$.
Proof of the idempotency relation reduces to the
orthogonality of the group character for the first case and
the unitarity of the modular transformation matrices 
for the second.
In both cases, the most essential property is that the structure constant is
written in Verlinde's form (\ref{Verlinde2},\ref{Verlinde}).

We compare these expressions with the Cardy states.  For the orbifold
 \cite{Billo:2000yb},
\begin{equation}
|\alpha\rangle =\frac{1}{\sqrt{|\Gamma|}}\sum_i 
\sqrt{\sigma_i}\,\zeta^{(\alpha)}_i
 |i\rrangle\,,\qquad |i\rrangle :=\sum_{g\in\cC_i} |g\rrangle\,.
\end{equation}
where $\sigma_i$ is a factor which appears in the
modular transformation of the character for the orbifold.
For the generic modular
character
$\chi_g^h(q)=\mbox{Tr}_{\mathcal{H}_g}(h \,q^{L_0-\frac{c}{24}})$,
we write their modular transformation as
$\chi_g^h(q)=\sigma(g,h)\chi_h^g(\tilde q)$. 
$\sigma_i$ is defined as
$\sigma(e,g)$ for $g\in\cC_i$.
For the RCFT, Cardy's solution is written as,
\begin{equation}
|\alpha \rangle=\sum_{i\in\mathrm{Primary}} 
\frac{S_{\alpha i}}{\sqrt{S_{1i}}} |i\rrangle\,.
\end{equation}
In either case, if we identify 
the base $e_i$ of group (resp. fusion) ring with
the normalized Ishibashi state 
$\sqrt{\sigma_i} |i\rrangle$ 
(resp. 
$(S_{i1})^{-1/2}|i\rrangle$
), Cardy states are the idempotents.

\paragraph{Discussion}
In this short note, we show that the Cardy states can be interpreted
as the idempotents of the group (fusion) ring.
It supports our conjecture \cite{Kishimoto:2003yu} that every consistent
boundary states
satisfy a universal nonlinear relation, the idempotency relation with
respect to the star product of the closed string field theory.
We note that  our discussion for RCFT can be formally extended to
the generic CFT with an infinite number of  primary fields.

We have to admit that our argument is so far weak from
the viewpoint of closed string field theory.
Namely we need to assume that the factorization
of the coefficients $\cC_{ij}{}^k(\alpha_1,\alpha_2)$ 
in Eq.(\ref{universal_rel})
without a detailed analysis of
the string vertex. 
For the generic CFT, the explicit proof is rather difficult
since there exists no oscillator definition 
of three string vertex for the generic background.
As we mentioned, however, 
an explicit formula
is known \cite{Itoh:1987aj} for the $\mathbf{Z}_2$ orbifold.
In our future publication, we will present our explicit evaluation of
the coupling constant and present a proof of our conjecture in this note.


We compare our discussion
with the more popular arguments of the description of D-branes
\cite{Gopakumar:2000zd}.
In a spirit that D-branes should 
be described by projectors, our discussion
is basically the same as theirs.  
However, while these arguments are based on the open string,
ours comes from the closed string. It induces some differences.
First, the algebra is non-commutative in open string approaches,
it is usually commutative in our discussion.
While the projectors in the open string theory pick up localized solitons
in the non-commutative space-time, ours specify the
representations that characterize the D-branes.   
The scope of these two approaches seems to be mutually complementary
but not contradictory.

\vskip 3mm
\noindent{\em Acknowlegement}: \hskip 3mm We are  obliged to
Y.~Nakayama, R.~Nobuyama, Y.~Tachikawa, H.~Takayanagi 
and E.~Watanabe
for helpful discussions.
I.~K. is supported in part by JSPS Research Fellowships for Young Scientists.
Y.~M. is supported in part by Grant-in-Aid (\# 13640267) from the
Ministry of Education, Science, Sports and Culture of Japan.

\end{document}